\documentclass{article}
\newcommand{\bfr}{\begin{flushright}}
\newcommand{\efr}{\end{flushright}}
 
\begin{document}
\title{Double Compactification
}
\author{ 
Atsushi Nakamula\\
Department of Physics, Tokyo Metropolitan University,\\
Setagaya-ku, Tokyo 158, Japan\\
and\\
Kiyoshi Shiraishi\\
Institute for Nuclear Study, University of Tokyo, \\
Midori-cho, Tanashi,
Tokyo 188, Japan
}
\date{Il Nuovo Cimento {\bf B105} (1990) pp.~179--190
}
\maketitle
\begin{abstract}
A cosmological scenario according to which our universe experienced
space-time compactifications twice in its early development is
investigated through toy models. In this scenario gauge configurations
on an extra space play essential roles to bring about a change of the
dimensionality of the compactified space. Simple models are offered and
their behaviour at finite temperature is examined. A possibility of
causing inflation and problems on our scenario is argued briefly. 
\end{abstract}

\section{Introduction}
During the last unified theories of forces, including gravity, in
higher-dimensional space-time have gained much interest. Promising
unification schemes such as supergravity \cite{1} and superstring
theories \cite{2} seem to make clear sense in higher dimensions. Usually
the extra dimensions, except for the ordinary space-time of four
dimensions, are assumed to be curled up in a tiny compact space \cite{3}.

In addition, the theories which contain fundamental gauge fields in
higher dimensions have been studied extensively \cite{3}. The theory of
this type may be deduced from superstring theories. In such theories, the
gauge field configuration on the extra spaces can play important roles
in compactification scheme. The energy and pressure of the gauge field
strengths on the compact space and the cosmological constant can lead to
stable compactification with a suitable balance. 

In this case, the
cosmological constant must be tuned by hand in order that our universe
in large dimensions results in a flat space-time. The same attitude is
taken in the usual inflationary universe scenario \cite{4}.

When we imagin the early era of the evolution of the universe, we can
presume the process of dynamical compactification. The dynamical
compactification has received much attention regarding its
cosmological aspects \cite{5}. There are various types of cosmological
scenarios with dynamical compactification which bring about inflationary
expansion \cite{6,7}.

In this paper, we show a new possible mechanism to serve the condition
for inflation in compactified theory with nontrivial configuration of
gauge field in extra dimensions. We think `phase transition' of the
gauge configuration. The phase transition is accompanied with the change
of the background geometry; even the dimensionality of compact subspace
changes. We wish to call the scenario `double compactification'. We
believe it worthwhile to analyse simple models on the basis of this
attractive possibility. It is shown that there appears a large
cosmological constant during the phase transition. The details and
subtleties are discussed later.

\section{Nontrivial configuration of gauge fields on an extra
space}
We take the Einstein-Yang-Mills Lagrangian
\begin{equation}
\frac{1}{\sqrt{-g}}L=-\frac{1}{2\kappa^2}R+\frac{1}{4e^2}{\rm tr}(F_{MN}F^{MN})+
\Lambda\,.
\label{eq1}
\end{equation}
Here $\kappa^2=8\pi G$; $G$ is the Newton constant; $e$ is a gauge coupling
constant; $\Lambda$ is a cosmological constant. The scalar curvature of $S^N$
with unit radius is defined as $R=+N(N-1)$.

We consider the $SU(2)$ gauge theory here. This may be regarded as a
subgroup of a large unified symmetry group.

Let us suppose the gauge fields on $S^3$ of unit radius with metric
\begin{equation}
d\Omega^2(S^3)=\tilde{g}_{mn}dy^mdy^n=d\psi^2+\sin^2\psi(d\theta^2+\sin^2
\theta d\phi^2)\,,
\label{eq2}
\end{equation}
where $0\le\psi, \theta<\pi$ and $0\le\phi<2\pi$.

We adopt the following ansatz in the matrix form of the gauge field
configuration:
\begin{eqnarray}
A_\psi&=&f\times\left(
\begin{array}{cc}
\cos\theta & \sin\theta\exp[-i\phi]\\
\sin\theta\exp[i\phi] & -\cos\theta
\end{array}\right)\,,\nonumber \\
A_\theta&=&f\times\sin\psi\cos\psi\left(
\begin{array}{cc}
-\sin\theta & \cos\theta\exp[-i\phi]\\
\cos\theta\exp[i\phi] & \sin\theta
\end{array}\right)\nonumber \\
& &+f\times\sin^2\psi\left(
\begin{array}{cc}
0 & -i\exp[-i\phi]\\
i\exp[i\phi] & 0
\end{array}\right)\,,\nonumber \\
A_\phi&=&f\times\sin^2\psi\sin\theta\left(
\begin{array}{cc}
\sin\theta & -\cos\theta\exp[-i\phi]\\
-\cos\theta\exp[i\phi] & -\sin\theta
\end{array}\right)\nonumber \\
& &+f\times\sin\psi\cos\psi\sin\theta\left(
\begin{array}{cc}
0 & -i\exp[-i\phi]\\
i\exp[i\phi] & 0
\end{array}\right)\,.
\label{eq3}
\end{eqnarray}

These configurations have been used in ref.~\cite{8} and in ref.~\cite{9}. For the
moment, we regard $f$ as a constant.

The energy density of the gauge field strength is
\begin{equation}
\frac{1}{4e^2}{\rm tr}(F_{MN}F^{MN})=\frac{12}{e^2}(f-f^2)^2\equiv V(f)
\quad
(\mbox{$f$ is a constant})\,,
\label{eq4}
\end{equation}
where the field strength is defined by
\begin{equation}
F_{mn}=\partial_mA_n-\partial_nA_m+i[A_m, A_n]\,.
\label{eq5}
\end{equation}

$A$ being a constant, the values of $f$ satisfying the field equation
\begin{equation}
D_mF^{mn}=\nabla_mF^{mn}+i[A_m, F^{mn}]=0\,,
\label{eq6}
\end{equation}
with $\nabla_m$ representing the covariant derivative on $S^N$, are $f=0$,
$1/2$ and $1$ \cite{9}. These values precisely correspond to the
stationary point of the potential. The gauge fields of $f=0$ and $f=1$
trivially satisfy the equation of motion (6) because they lead vanishing
field strengths. The gauge configuration (3) with $f=1$ turns out to be
locally gauge-equivalent to the trivial configuration $f=0$ \cite{8}.
Later on, we will pay attention to the configuration with $f=1/2$.

Incidentally, this configuration may remind us of the physics of
``sphaleron'' \cite{10}. But the configuration considered here exists in
pure Yang-Mills theory and can have finite energy only in a compact
space. Now let us get back to our subject. We shall work in the
$(d+1+3)$-dimensional space-time with metric
\begin{equation}
ds^2 =ds^2(d+1)+b^2d\Omega^2(S^3)\,,
\label{eq7}
\end{equation}
and we assume the $(d+1)$-dimensional space-time admits the metric of
an Einstein space. First we suppose the previous gauge field (\ref{eq3}) with $f=
const$ on the extra space.

The substitution of the above configuration into the Einstein equations
following from (\ref{eq1}) leads to
\begin{eqnarray}
R_{\mu\nu}^{(d+1)}&=&\frac{\kappa^2}{d+2}\left(2\Lambda-\frac{24(f-f^2)^2}{e^2b^4}
\right)g_{\mu\nu}^{(d+1)}\,,
\label{eq8a}\\
R_{mn}&=&\frac{\kappa^2}{d+2}\left(2\Lambda+\frac{8(2d+1)(f-f^2)^2}{e^2b^4}
\right)g_{mn}\,,
\label{eq8b}
\end{eqnarray}
where $R_{\mu\nu}^{(d+1)}$ is a Ricci tensor derived from the metric
$g_{\mu\nu}^{(d+1)}$ of the $(d+1)$-dimensional Einstein space, while $R_{mn}$ is a
Ricci tensor for the compactified space and $g_{mn}=b^2\tilde{g}_{mn}$.

In the cases with $f=0$ and $f=1$, eqs.~(\ref{eq8a}, \ref{eq8b}) have a static
solution in which the extra-space is expressed as a static sphere with radius 
$\{(d+1)/\kappa^2\Lambda\}^{1/2}$ and the
$(d+1)$-dimensional space-time looks like the de Sitter space. However, this
solution is unstable; the extra-space will collapse $(b\rightarrow 0)$ or will
become large indefinitely $(b\rightarrow\infty)$.

In this case with $f=1/2$, (\ref{eq8b}) has a static and stable solution for
appropriate values of $\Lambda$, $\kappa^2$ and $e^2$ and the $(d+1)$-dimensional
space-time becomes de Sitter space.

The preferred value of the cosmological constant will be decided in the next
section. We suspend an analysis of the solution of (\ref{eq7}) until the
examination of the classically stable structure of the gauge field.

\section{Compactification caused by a domain wall}
In this section we consider $f$ in (\ref{eq3}) as a function of one of the space
coordinates, say $z$. In a flat-space background, the existence of a
solitonlike structure of the gauge field has been found \cite{8}. Here, we
take the curved background geometry and Einstein equations into account.
To this end, we consider the following metric in $(d+1+3)$-dimensional
space-time: 
\begin{equation}
ds^2=ds^2(d)+a(z)^2dz^2+b(z)^2d\Omega^2(S^3)\,.
\label{eq9}
\end{equation}
with $d$-dimensional space-time of the Einstein metric and $a(z)$ and $b(z)$ are
functions of $z$. The field equation for the Yang-Mills field (\ref{eq6}) is
reduced to
\begin{equation}
(a^{-1}bf')'=4ab^{-1}(f-f^2)(1-2f)\,,
\label{eq10}
\end{equation}
where $'$ denotes the derivative with respect to $z$. To solve this equation, we
take an ansatz
\begin{equation}
a(z)\propto b(z)\,.
\label{eq11}
\end{equation}

Then eq.~(\ref{eq10}) becomes
\begin{equation}
f ''=4(a/b)^2(f-f^2)(1-2f)\,,
\label{eq12}
\end{equation}
where $(a/b)$ is a constant. This equation is the same as that in flat space, up to
inclusion of an undetermined constant $(a/b)$. For this case, we have the kink
solution, which exists in the $\phi^4$-theory \cite{8,11}:
\begin{equation}
f(z)=\frac{1}{2}\left(1+\tanh\left(\frac{a}{b}z\right)\right)\,.
\label{eq13}
\end{equation}

We need the Einstein equations to determine the function $a(z)$ or $b(z)$. The
equations are reduced to
\begin{eqnarray}
R_{\mu\nu}^{(d)}&=&\frac{\kappa^2}{d+2}\left\{2\Lambda-\frac{1}{e^2}
\left(\frac{6(f')^2}{a^2b^2}+\frac{24(f-f^2)^2}{b^4}\right)\right\}g_{\mu\nu}^{(d)}\,,
\label{eq14a}\\
R_{zz}&=&\left(-\frac{3}{a}\left(\frac{b'}{ab}\right)'-3
\left(\frac{b'}{ab}\right)^2\right)g_{zz}\nonumber \\
&=&\frac{\kappa^2}{d+2}\left\{2\Lambda+\frac{1}{e^2}
\left(\frac{6(d+1)(f')^2}{a^2b^2}-\frac{24(f-f^2)^2}{b^4}\right)\right\}g_{zz}\,,
\label{eq14b}\\
R_{mn}&=&\left(-\frac{1}{a}\left(\frac{b'}{ab}\right)'-3\left(\frac{b'}{ab}\right)^2
+\frac{N-1}{b^2}\right)g_{mn}\nonumber \\
&=&\frac{\kappa^2}{d+2}\left\{2\Lambda+\frac{1}{e^2}
\left(\frac{2(d-1)(f')^2}{a^2b^2}+\frac{8(2d+1)(f-f^2)^2}{b^4}\right)\right\}g_{mn}\,,
\label{eq14c}
\end{eqnarray}
where $R_{\mu\nu}^{(d)}$ is a Ricci tensor derived from the metric
$g_{\mu\nu}^{(d)}$ of
$d$-dimensional space-time and $R_{zz}$ is a component of the Ricci tensor and 
$g_{zz}=a^2(z)$. In addition, we shall look for solutions such that
\begin{equation}
R_{\mu\nu}^{(d)}=0\,,
\label{eq15}
\end{equation}
i.e. we demand flat large dimensions. By substituting the kink solution of $f(z)$
and flat $d$-dimensional metric, we find the following solution:
\begin{eqnarray}
a(z)&=&A\{\cosh(A/B)z\}^{-1}\,,
\label{eq16a}\\
b(z)&=&B\{\cosh(A/B)z\}^{-1}
\label{eq16b}
\end{eqnarray}
with
\begin{equation}
B^2=\kappa^2/2e^2\,.
\label{eq16c}
\end{equation}
(Here $A$ is an arbitrary constant which defines the unit of length scale in
$z$-direction) and algebraic relations between $\kappa^2$, $e^2$ and $\Lambda$:
\begin{equation}
\Lambda=\frac{6e^2}{\kappa^4}\,.
\label{eq17}
\end{equation}
(Note that eqs.~(\ref{eq16a}, \ref{eq16b}, \ref{eq16c}) and (\ref{eq17}) are
independent of the dimensionality
$d$.) Does this space-time realize the `universe in a domain-wall' suggested by
Rubakov and Shaposhnikov \cite{12}? Unlike their consideration, our solution is
coupled with gravity. However, a close examination reveals another aspect of
compactification.

Let us try to rewrite the solution using a new coordinate as
\begin{equation}
r=C\exp [(A/B) z]\,,
\label{eq18}
\end{equation}
where $C$ is a constant.

Performing this substitution in the solution above, we get the metric
\begin{equation}
ds^2=ds^2(d)+\frac{4C^2B^2}{(C^2+r^2)^2}\{dr^2+r^2d\Omega^2(S^3)\}\,.
\label{eq19}
\end{equation}

Thus this turns out to be a direct-product space $M_d \times S^4$ ($d$-dimensional
Minkowski space-time$\times$four-dimensional sphere). (If one considers another
variable $y$ defined as $\sin y =\tanh (A/B)z$, then one obtains $ds^2=ds^2(d)+
B^2\{dy^2+\cos^2 y d\Omega^2(S^3)\}$.)

If we transform the coordinates into the solution of gauge structure, we find the
coincidence with the solutions considered in ref.~\cite{13}, dubbed as
`instanton-induced compactification'. Therefore, fortunately, the stability of
the solution has been guaranteed. Moreover, it is known that there can be massless
fermion field coupled to the gauge field and the model can be taken as a
quasi-realistic model. 

Substituting the algebraic relation back into the
eqs.~(\ref{eq8a}), (\ref{eq8b}), we find the $(d+1)$-dimensional space-time is a
de Sitter space as long as eq.~(\ref{eq8b}) has static solutions. This point is
discussed again later.

\section{Cosmological scenarios}
Now we are ready to state a new cosmological scenario. Suppose that, once
upon a time, we lived in $(d+1)$-dimensional space-time with the compact space
$S s$ and the gauge configuration (\ref{eq3}) with $f=1/2$ is realized by some
reason (see later). During or after the de Sitter expansion, the universe was
divided into domains in which the gauge configuration on $S^3$ is expressed as the
form of (\ref{eq2}) with $f=0$ or $f=1$, i.e. the potential minima. Then the domain
wall between the domains of $f=0$ and of $f=1$ appeared. Once the domain wall was
made, the compactification of the perpendicular direction to the wall occurred.
Then, if we were in the domain wall, we lived in flat $d$-dimensional space-time
with the compact space $S^4$, provided the algebraic relation between the coupling
constants holds. Note also that we can consider the gauge symmetry breaking
caused by the nontrivial gauge configuration.

Here we must suppose the size of $S^3$ is contracting $(b\rightarrow 0)$ during the
phase transition in the region of $f= 0$ and $f=1$, because of the form of the
solution (\ref{eq16a}, \ref{eq16b}, \ref{eq16c}). The domain-wall structure
between the regions in which the size of the extra-space expands indefinitely is
presently unknown. This will be clarified in future work. 

The universe filled with
infinite domains was considered by Linde
\cite{14}. He considered that even the dimensionality of space-time was different
from domain to domain \cite{15}. But in our scenario, we should live in one
domain-wall, not in a domain.

Now, we must consider the realization of the initial configuration of gauge
fields. We have two ideas at present. One of the ideas is that we consider a
radiation-dominated era before the `inflation' as in the `old and new'
inflationary cosmology scenario \cite{4}. To explain this feasibility, we propose a
simpler model for our ``double compactification'' scenario.

Let us consider a two-dimensional sphere with standard polar coordinates.
We consider an $SU(2)$ gauge structure of the following type:
\begin{eqnarray}
A_\theta&=&f\times \left(\begin{array}{cc}
0& -i\exp[-i\phi]\\
i\exp[i\phi]&0
\end{array}\right)\,,\label{eq20a}\\
A_\phi&=&f\times\sin\theta\left(\begin{array}{cc}
\sin\theta & -\cos\theta\exp[-i\phi]\\
-\cos\theta\exp[i\phi]&-\sin\theta
\end{array}\right)\,.
\label{eq20b}
\end{eqnarray}

One can find these gauge fields yield similar potential energy as in the
previous case with an appropriate scale in a vertical line. Therefore we can
obtain a solitonic solution or a domain-wall structure in this simple model, though
``topological'' stability is not garanteed unlike the previous $S^3$ case (see
ref.~\cite{8}). The configuration satisfies the general spherical symmetric ansatz
which yields spherically symmetric distributions to the energy density, like the
previous case. 

When $f=1/2$, the configuration is gauge-equivalent to the monopole
configuration \cite{16,17}. To see this, we take the following gauge
transformation: 
\begin{equation}
A_m\rightarrow A'_m=\Omega A_m\Omega^\dagger-i\Omega\partial_m\Omega^\dagger\,,
\label{eq21a}
\end{equation}
with
\begin{equation}
\Omega=\left(\begin{array}{cc}
\cos(\theta/2)&\sin(\theta/2)\exp[-i\phi]\\
-\sin(\theta/2)\exp[i\phi] &\cos(\theta/2)
\end{array}\right)\,.
\label{eq21b}
\end{equation}
Then we get, provided $f=1/2$,
\begin{eqnarray}
A'_\theta&=&0\,,\\
\label{eq22a}
A'_\phi&=&\frac{1}{2}\left(\begin{array}{cc}
1-\cos\theta&0\\
0& -(1-\cos\theta) 
\end{array}\right)\,,
\label{eq22b}
\end{eqnarray}

The classical instability at this configuration was discussed in ref.~\cite{17},
and the quantum correction to vacuum energy was investigated by Hosotani in ref.
\cite{16}; Note that gauge symmetry is broken at this point, $f=1/2$ \cite{16}.

We examine the free energy at finite temperature in addition to the potential
energy at the typical points with $f=0$ and $f=1/2$. We adopt $SU(2)$-doublet
fermions as matter fields. The one-loop free energy of the massless fermion gas
at temperature $T$ in the background space-time $M_{d+1} \times S^2$ and no gauge-
background are formally given by
\begin{equation}
F(0)=\frac{4N_FV_d\cdot 2^{[(d+1)/2]}}{\beta(4\pi)^{d/2}}\sum_{n=-\infty}^\infty
\sum_{l=0}^\infty(l+1)\left(\frac{(l+1)^2}{b^2}+\left(\frac{2\pi}{\beta}\right)^2(n+1/2)^2\right)^{d/2}\,,
\label{eq23}
\end{equation}
where $\beta=T^{-1}$ and $N_F$ is the number of fermion species. $V_d$ is the
$d$-dimensional volume of the system.

Here we consider the large dimensions as a flat space, because before the
``phase transition'' the universe is assumed to be dominated by radiation and
appears to be a flat space approximately as in ordinary inflationary scenarios.
The free energy of the fermions coupled to the gauge configuration (\ref{eq20a},
\ref{eq20b}) with
$f=1/2$ is expressed as
\begin{equation}
F(1/2)=\frac{N_FV_d\cdot 2^{[(d+1)/2]}}{\beta(4\pi)^{d/2}}\sum_{n=-\infty}^\infty
\sum_{l=0}^\infty
d(l)\left[\frac{l(l+1)}{b^2}+\left(\frac{2\pi}{\beta}\right)^2(n+1/2)^2\right]^{d/2}\,,
\label{eq24}
\end{equation}
where $d(l)=2(2l+1)$ when $l>0$ and $d(0)=1$.

At high temperature $(T\gg 1/b)$, the free energy in each case is approximately
given by
\begin{eqnarray}
& &F(0)\sim-\frac{4N_F(4\pi b^2)V_d\cdot 2^{[(d+1)/2]}}{(4\pi)^{(d+3)/2}}\Gamma
\left(\frac{d+3}{2}\right)\nonumber \\
&
&\quad\times\left[\left(1-\frac{1}{2^{d+2}}\right)\zeta(d+3)\frac{2^{d+3}}{\beta^{d+3}}
-\frac{1}{3(d+1)}\left(1-\frac{1}{2^{d}}\right)\zeta(d+1)
\frac{2^{d+1}}{b^2\beta^{d+1}}\right]\,,\\
\label{eq25a}
& &F(1/2)\sim-\frac{4N_F(4\pi b^2)V_d\cdot 2^{[(d+1)/2]}}{(4\pi)^{(d+3)/2}}\Gamma
\left(\frac{d+3}{2}\right)\nonumber \\
&
&\quad\times\left[\left(1-\frac{1}{2^{d+2}}\right)\zeta(d+3)\frac{2^{d+3}}{\beta^{d+3}}
+\frac{2}{3(d+1)}\left(1-\frac{1}{2^{d}}\right)\zeta(d+1)
\frac{2^{d+1}}{b^2\beta^{d+1}}\right]\,.
\label{eq25b}
\end{eqnarray}

Thus if $F(0)-F(1/2)>1/4e^2b^4\times V_d\times(4\pi b^2)$, the difference of
one-loop free energy overcomes the difference of the tree level potential energy.
In other words, at sufficiently large $T~(>N_F(e^2b^2)^{-1/(d+1)})$, the vacuum
expectation value of
$f$ is $1/2$ in the presence of fermion doublets  If a sufficiently large number
of fermions exist, their contribution overwhelm the radiation of gauge particles
and other matter contents in the system under consideration.

This result supports the following speculation: in a very early universe, it is
very hot and filled with radiations of matter fields. In a later era in the history of
the universe, the vacuum value of $\langle f\rangle$ cannot tolerate its initial
value
$f=1/2$ as temperature decreases. Then there appear many domains and domain walls;
``our universe'' should be contained in one of the domain wall. During this
transition inflation may take place.

The previous model with extra-space of $S^3$ is expected to have a similar
property at finite temperature; phase transition of gauge configuration could
take place.

The point $f=1/2$ might not be global minimum but local minimum of the
potential. In any case, the domain wall is, at least, unstable. The precise shape of
the effective potential is crucial for cosmological evolution as in the inflationary
universe scenario.

Another attractive scenario works in a cold universe. At first our universe has
$(d+1)$ large dimensions and $S^3$ (or $S^2$) as a compact space. Suppose that the
matter field contribution to energy density is negligible. In the early stage of
cosmological evolution, the $(d+1)$-dimensional space can be de Sitter or de
Sitter-like space regardless of the value of $f$. In other words, we assume the
extra-sphere had not yet collapsed nor been decompactified. On the other hand,
the effective action of $f$ is precisely the same as that of a scalar field with
self-couplings. In the de Sitter space, it is known that the quantum fluctuation
drives the vacuum value of scalar fields \cite{18}. This dynamics has been studied
by many authors through a Fokker-Planck-type equation, and often called stochastic
dynamics \cite{19}. Thus the vacuum value o f f can cross over the potential
barrier between $f=0$ and $f=1$; stochastic processes could account for the
completion of such a transition. Consequently it is possible to make many domains
in which the gauge configuration takes $f=0$ or $f=1$.

This is a ``cold'' universe scenario. It seems very interesting to investigate the
development of $f$ in this scenario using computer simulations.

\section{Comments}
Now several comments are in order. First, in this paper, we tacitly
assume the static extra-space in its size before the ``phase
transition'' as in the arguments of Wilson-loop symmetry breaking \cite{20}.
At the classical level, i.e. as a solution of (\ref{eq8b}), we can obtain static
extra-spheres if $d\ge 9$ in our former model with $S^3$ and $d\ge 10$ in our
latter model with $S^2$. (In each case, $(d+1)$-dimensional space-time
becomes de Sitter space of which the distance of horizon is of order
$\sim(e^2/\kappa^2)^{-1/2}\sim(\kappa^2\Lambda)^{-1/2}$). Otherwise, if we want a
static compact space before the phase transition when $d<9$, we obtain anti-de
Sitter space after the compactification of another dimension. If this is true, we
must consider this model as an intermediate stage of the history of the
universe, i.e. successive compactifications occur until the large
dimensions left become four.

However, we must take quantum corrections from matter and gravitational
fields into account. Furthermore, even at the classical level, higher-order terms
in the curvatures in the action can save the stability of the extra space. These
possibilities will be examined in our subsequent work \cite{21}.

Secondly, we looked for de Sitter solutions of gravitational equations of motion
to test the possibility of the inflationary scenario. There is a subtlety, however, in
the occurrence of inflation, because of the simultaneous change in spatial
dimensionality. Moreover, the model with $d<9$ has a possibility in serving
another dynamical evolution including inflating large spaces. Note that an
inflationary universe scenario does not require an ``exact'' de Sitter space. The
study of dynamical evolution of the model along with the cosmic time is necessary.

Several initial-condition problems are known in inflation in Kaluza-Klein
theories. In one of the scenarios, a selected initial condition for scale factors
or background geometry is crucial \cite{6}. Another scenario makes use of the
dynamical evolution of scale factors of the compact space as a source of the rapid
expansion in large dimensions \cite{7}. But even in this type of scenario, the
compatibility with the stable compactification after inflation requires
a restriction to the initial condition in the scale factors of large and
small dimensions \cite{22}.

Though our scenario of ``double compactiflcation'' is very different from the
above mechanisms, the initial condition of our scenario for inflation may be
restricted within some range of variable quantities. We must carefully
investigate dynamical evolution of double compactification. We would like to
suggest that a numerical simulation like in ref.~\cite{23} may be effective, for
example, in our ``hot'' universe scenario.

Thirdly, if we consider the ``hot'' scenario at finite temperature, we should
consider finite-temperature instability for compactification \cite{24}. The quantum
effects in curved space and self-consistent background in terms of Einstein
equations have to be considered \cite{25}. Moreover, if we wish to
investigate our models closely, we must consider dynamical quantum
effects \cite{26} too. Even thermal effects in a nonequilibrium system
ought to be studied. Together with the previous problem of spatial
dimensionality admitting stable compactiflcation, the aspects of quantum
and thermal effects will be studied extensively \cite{27}. 

There are many
other topics; the effect of $F^4$ term or higher-order terms in field
strengths in the action suggested in the string theory; a search for
nontrivial gauge configurations in other gauge groups on complicated
compact spaces; the nontrivial structure of gauge fields on a flat
compact space, such as a torus, which may have relation with the string
theory; the reheating mechanism after the phase transition; the universe
before the double compactification or ``triple'' compactification which
explains the origin of gauge field and compact spaces. Each topic is an
interesting subject of much worth to study.

Finally we mention the cosmological constant problem \cite{28}. Coleman
suggested a possible solution to the cosmological constant in
four-dimensions in the framework of quantum gravity \cite{29}. In his
scenario even the fundamental couplings behave like dynamical variables
through quantum-gravity effects. If we believe this scenario, we must
re-examine our model; the algebraic relation to admit a flat-space
solution might become an ``equation of motion'' of coupling ``constants'',
$\Lambda$, $\kappa^2$ and $e^2$.

Several authors made efforts to describe other theories of a Kaluza-Klein type
in the context of quantum gravity \cite{30}. We will report on this topic
in relation to our model.

In this paper, we outlined the double-compactification scenario which
provides the feasibility of causing inflationary expansion in our large dimensions.
Many subjects to study further are left for future work. We must start from the
investigation of this new scenario from various points of view.

\bigskip

\bigskip

This work is supported in part by the Grant-in-Aid for Encouragement of
Young-Scientist from the Ministry of Education, Science and Culture
(\# 63790150).
One of the authors (KS) is grateful to the Japan Society for the Promotion of
Science for the fellowship. He also thanks Iwanami F\=ujukai for the financial
aid.


\end{document}